# Impact of High PV Penetration on Regional Power Grids

Shutang You
University of Tennessee, Knoxville, TN, USA
Email: syou3@utk.edu

*Abstract*— Due to the high solar irradiance or energy price, certain regions in the U.S. may reach 100% PV penetration and experience degradation of frequency response greater than the interconnection as a whole. Therefore, in this section, the 100% PV penetration region in each interconnection is simulated to study the local high PV penetration effects. The study was performed by quantifying RoCoF, frequency nadir, and settling frequency at different regional PV penetration levels. The impact of high regional PV penetration on the compliance of grid code on frequency response is also studied.

## I. Introduction

Ensuring power grid reliability is important to the society and economy. With the increase of renewable penetration, the characteristics of the power grid is changing. Existing wide-area monitoring systems deployed in power grids have also noticed the impact of high renewable penetration from both the interconnection level and the local level. Some previous studies have focused on the degradation of system frequency response as a whole system. Other studies focuses on the impacts of high renewable penetration on other aspects, such as voltage stability, oscillation, transient stability, etc. Some methods to mitigate the inter-connection level impacts have also been developed. As the influence of high renewable penetration on the interconnection level is better understood, the knowledge of how high renewable penetration in the regional power grids is still insufficient.

In this paper, the impact of regional high PV penetration is studied by quantifying the RoCoF, frequency nadir, and setting frequency at different regional PV penetration levels. This study is based on high renewable study models of the U.S. Eastern Interconnection (EI) model and ERCOT model. In addition, the grid code compliance under high regional PV penetration is also studied to provide some recommendations to future frequency response grid code revision.

## II. 4.1 Impact of 100% PV Penetration on the Regional Level Frequency Response

### A. Impact of 100% PV Penetration - EI Case Study

In PV distribution results, PV penetration of the PJM_ROM region in the EI may reach 100% due to its high energy price. To investigate the impact of 100% PV penetration, the largest contingency recommended by NERC (4.5 GW generation loss) was simulated inside this region. The locations of PJM_ROM region and the largest contingency are shown in Figure 1.

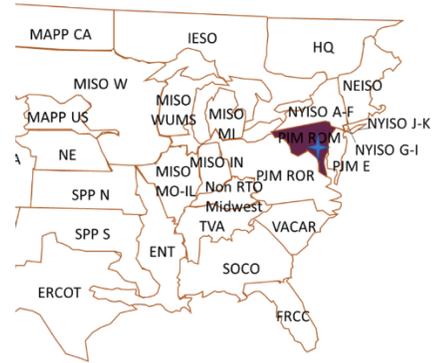

Figure 1. The 100% PV penetration region (PJM_ROM) and the contingency location (indicated by the blue star) in the EI

Figure 2 shows the average frequency responses of PJM_ROM region and the entire EI after the 4.5 GW generation loss contingency in each scenario while Figure 3 presents the corresponding frequency response metrics. These results demonstrate that despite the obvious oscillations in the regional frequency response, the overall frequency response trend remains the same at both the regional and interconnection levels. This is mainly because PJM_ROM has strong connections with its neighboring regions so its frequency can be supported quickly by its neighbors. As shown in Figure 2, PJM_ROM region has a sharper frequency decline right after the generation loss due to the local oscillations. Therefore, a larger ROCOF can be observed for the PJM_ROM region in Figure 3. However, as these oscillations gradually damped out, no major differences can be noticed in terms of frequency nadir and settling frequency. These observations indicate that the 100% PV penetration in PJM_ROM will not cause major operation difficulties for this region in terms of frequency response.

This work was also supported by U.S. Department of Energy Solar Energy Technologies Office under award number 30844. This work made use of Engineering Research Center shared facilities supported by the Engineering Research Center Program of the National Science Foundation and the Department of Energy under NSF Award Number EEC-1041877 and the CURENT Industry Partnership Program.

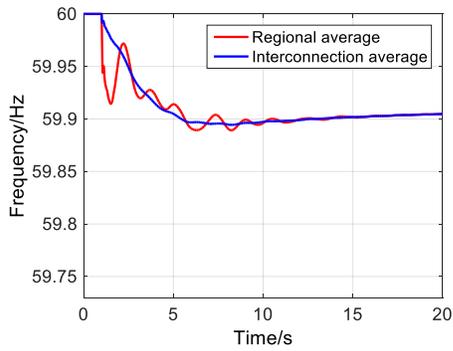
(a) 20% renewable at the interconnection level

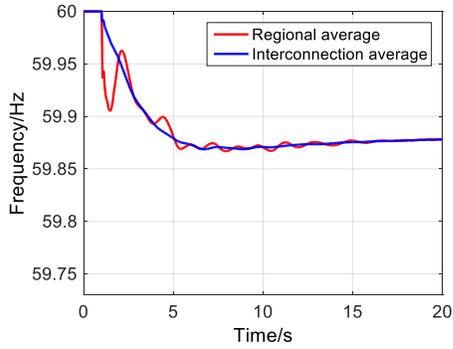
(b) 40% renewable at the interconnection level

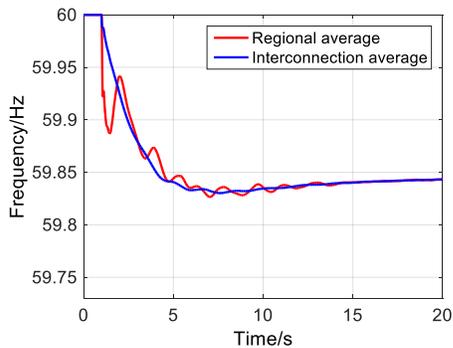
(c) 60% renewable at the interconnection level

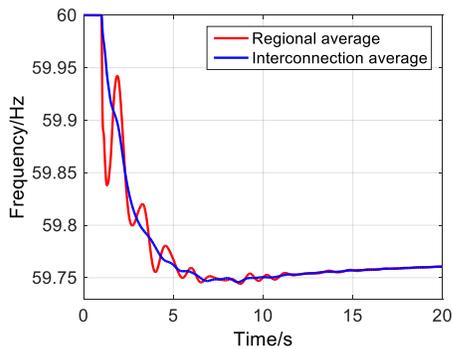
(d) 80% renewable at the interconnection level

Figure 2. EI 100% PV penetration region and interconnection frequency response (4.5 GW contingency)

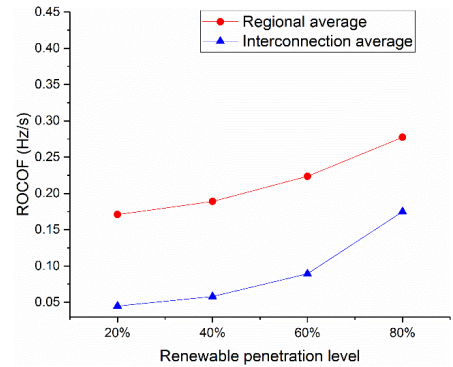
(a) ROCOF

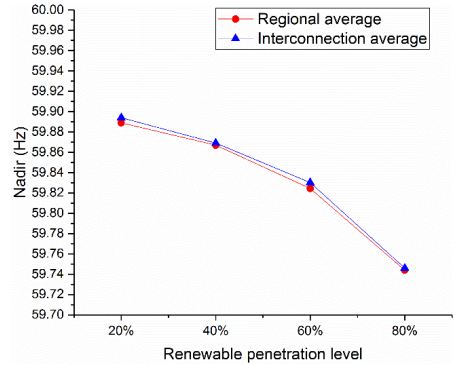
(b) Frequency nadir

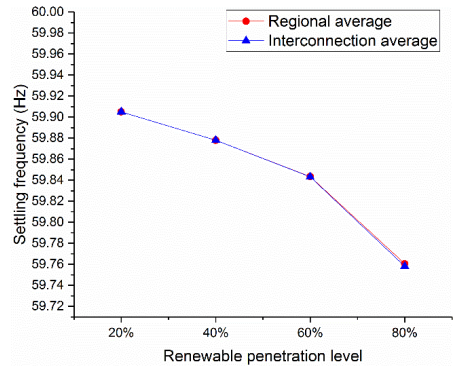
(c) Settling frequency

Figure 3. EI 100% PV penetration region and interconnection frequency response metrics (4.5 GW contingency)

As required by the SOPO document, four contingencies with the same generation loss magnitude were simulated at different locations within the PJM_ROM region to quantify the prediction uncertainties of the regional frequency response metrics. Figure 4 shows the locations of these four contingencies and Figure 5 shows their frequency response metrics. As demonstrated by Figure 5, the prediction uncertainties of all frequency response metrics meet the SOPO requirements (within ±5% of the mean value for N fixed magnitude events at different locations), which means the location of contingency does not influence frequency response metrics significantly.

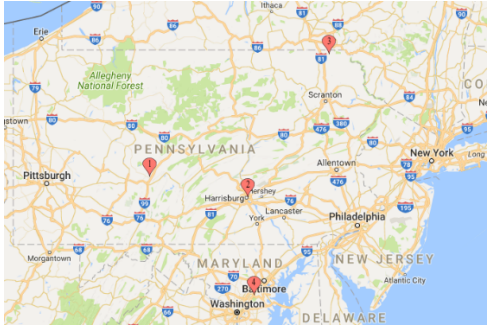

Figure 4. Contingency locations in the EI 100% PV penetration region (PJM_ROM)

## B. Impact of 100% PV Penetration - ERCOT Case Study

In PV distribution results, the ERCOT system is expected to have a 100% PV penetration region in the Austin-Greater San Antonio (A-GSA) area with a total generation of 11GW (shown in Figure 6). A largest N-2 generation loss contingency (2.75 GW) was simulated inside this region and the average frequency response curves of the A-GSA region and the entire ERCOT are given in Figure 7. Furthermore, the frequency response metrics such as ROCOF, frequency nadir, time to reach the nadir, and settling frequency, are presented in Figure 8. For each of these metrics, both the regional average value and the interconnection average were calculated.

The results from Figure 7 and Figure 8 demonstrate that there is hardly any difference between the regional average and interconnection average frequency responses. There exists a slight difference for ROCOF for lower levels of renewable penetration in Figure 8. This is due to small oscillations within the regional average frequency that causes some discrepancy when calculating these values. These oscillations occur most noticeably right after the event and begin subsiding at the nadir. Note that these oscillations are so small that it is difficult to see them in Figure 7 without zooming in.

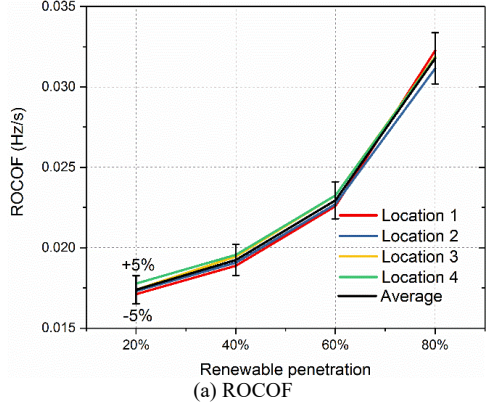
(a) ROCOF

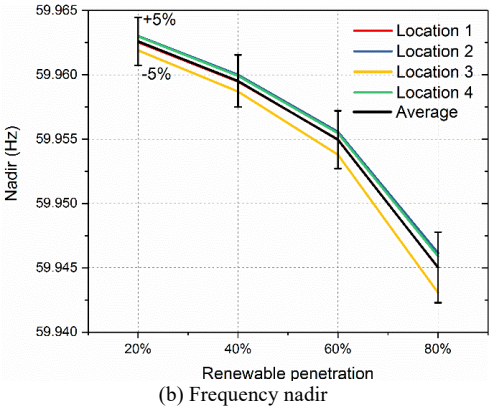
(b) Frequency nadir

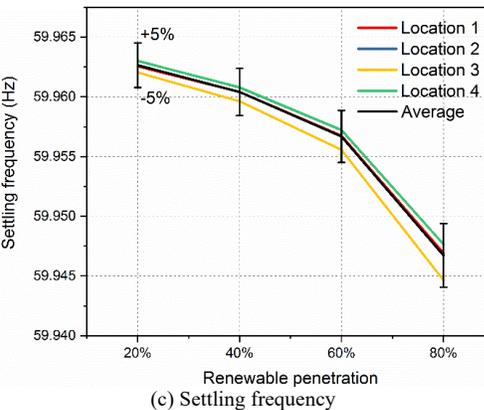
(c) Settling frequency

Figure 5. EI regional frequency response metrics for similar events at different locations

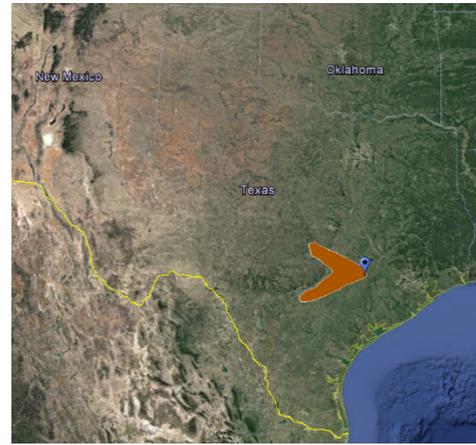

Figure 6. 100% PV penetration region (Austin-Greater San Antonio) and contingency location in the ERCOT

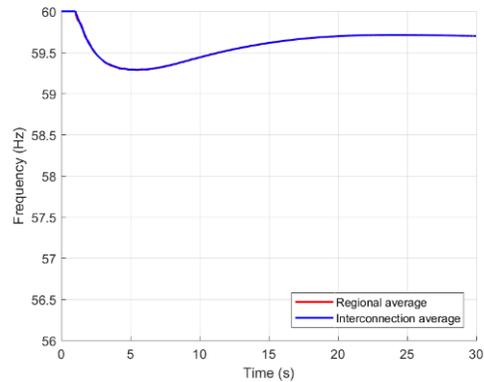

20% renewable at the interconnection level

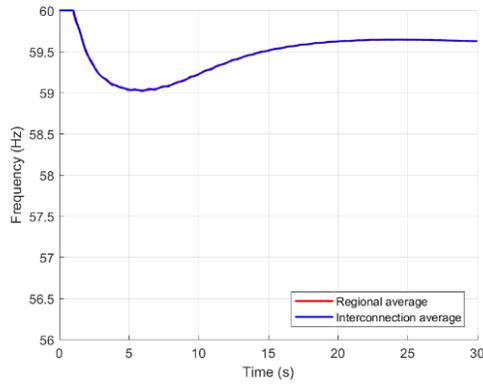

(a) 40% renewable at the interconnection level

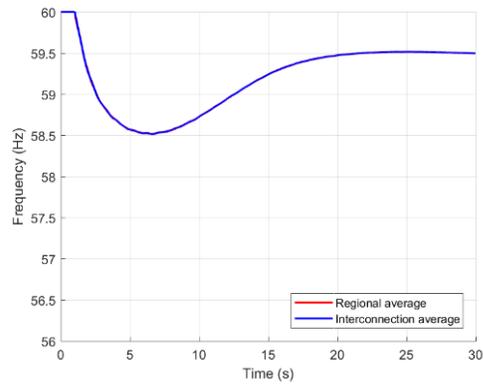

(c) 60% renewable at the interconnection level

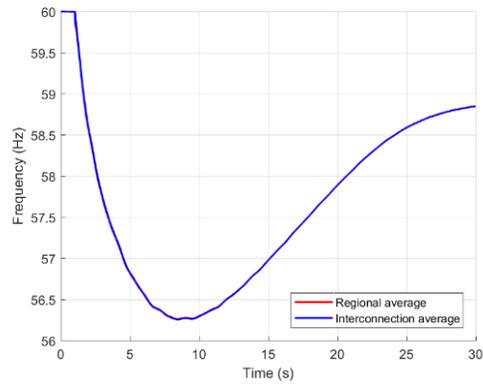

(d) 80% renewable at the interconnection level

Figure 7. ERCOT 100% PV penetration region and interconnection frequency response (2.75 GW generation loss contingency)

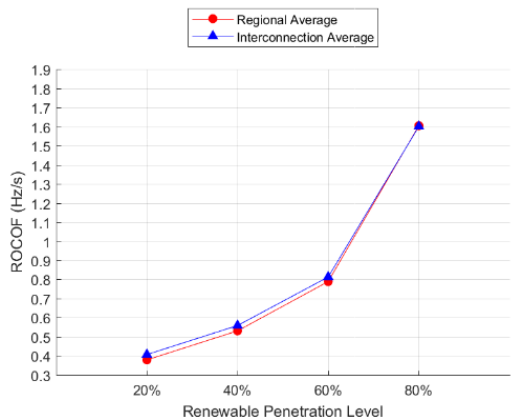

(a) ROCOF

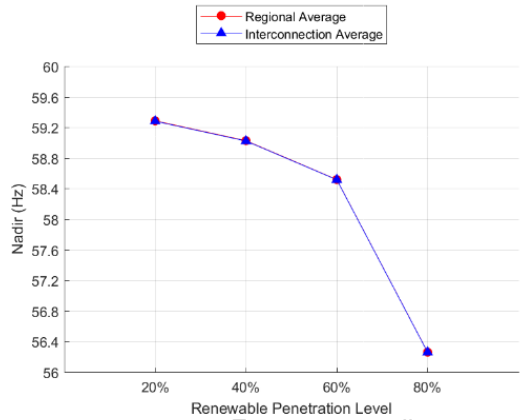

(b) Frequency nadir

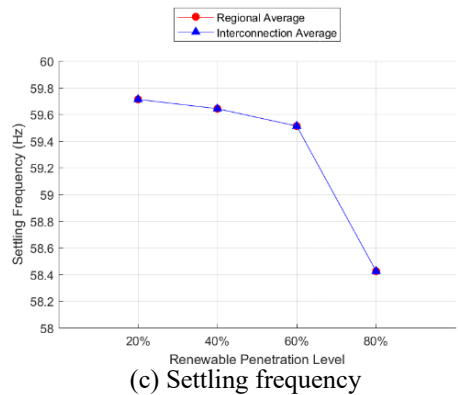

(c) Settling frequency

Figure 8. ERCOT regional frequency response metrics for similar events at different locations

In addition, three contingencies with the same generation loss amount within the A-GSA region were used to determine the prediction uncertainty of the regional frequency response metrics. Figure 9 shows the locations of the three similar events and Figure 10 shows the results of the simulations. Upon close inspection, all of these results fall within the 95% confidence interval, which fulfills the SOPO requirement. The results show that the difference in event location had little or no impact on any of the frequency response metrics.

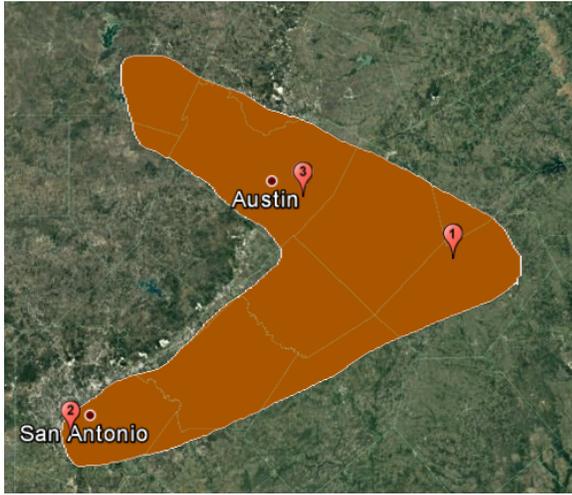

Figure 9. Contingency locations in 100% PV regional area in ERCOT

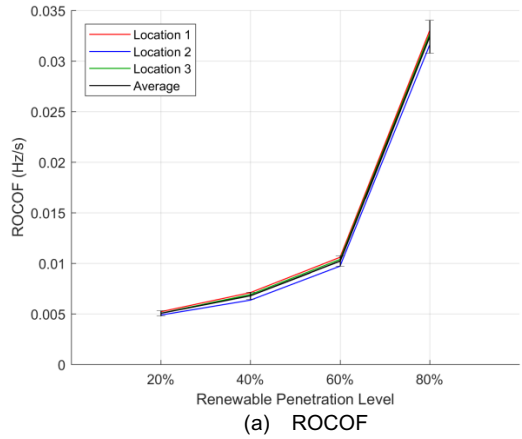

(a) ROCOF

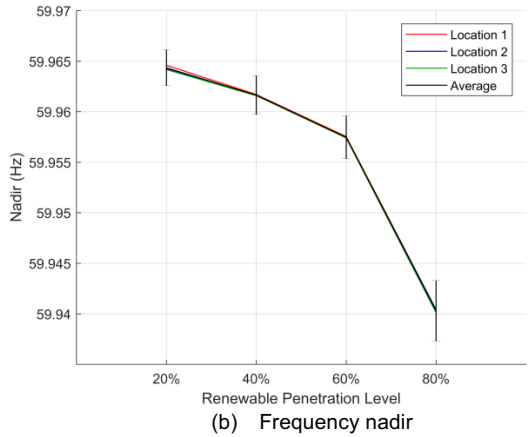

(b) Frequency nadir

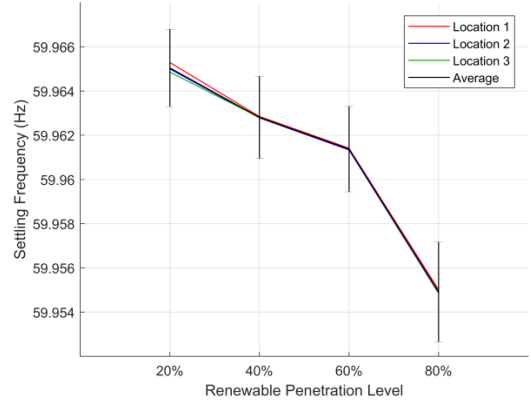

(c) Settling frequency

Figure 10. ERCOT regional frequency response metrics change for similar events

### III. IMPACT OF HIGH PV PENETRATION ON THE REGIONAL GRID CODE COMPLIANCE

Balancing authorities (BAs) are responsible for balancing power system demand and supply in real time. This study assesses the impact of high PV penetration on the compliance of the frequency response grid code at the BA level in the EI. The interconnection-level frequency response obligation is allocated to each BA according to Equation (1) [1].

$$\text{FRO}_{BA} = \text{IFRO} \times \frac{\text{Annual Gen}_{BA} + \text{Annual Load}_{BA}}{\text{Annual Gen}_{Int} + \text{Annual Load}_{Int}} \quad (1)$$

where IFRO is the interconnection frequency response obligation; Annual Gen$_{BA}$ and Annual Load$_{BA}$ are respectively the total annual electricity generation and consumption within each BA; Annual Gen$_{Int}$ and Annual Load$_{Int}$ are respectively the total annual generation and consumption of the entire interconnection.

The actual frequency response of BA $i$ is calculated by:

$$\text{RFR}_i = \Delta P_i / \Delta f \quad (2)$$

where $\Delta P_i$ is the real power output difference between Point A and Point B (in NERC specification [2]) for BA $i$; while $\Delta f$ is the frequency difference between Point A and Point B.

In the EI, the contingency for regional frequency response assessment is the largest resource event in last 10 years (4.5 GW generation loss) [1]. The frequency responses of various BAs were calculated using Equation (2) in each scenario. Figure 11 and Figure 12 show the major BAs' frequency response values and renewable penetration rates in different PV penetration scenarios.

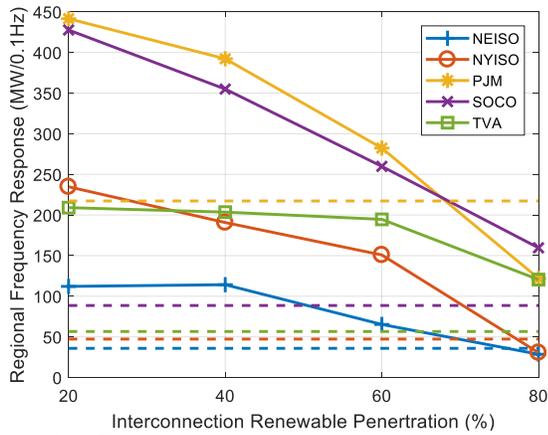

Figure 11. Change of regional frequency response in the EI
(Dash lines represent **FRO$_{BA}$**[1])

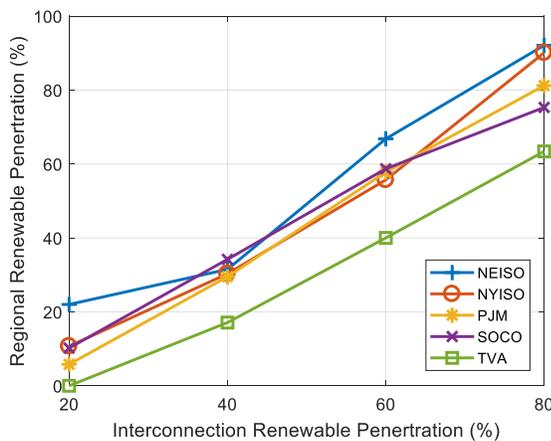

Figure 12. Change of regional renewable penetration in the EI

There is no doubt that the regional frequency response decreases as PV penetration increases. Specifically, PJM, NYISO, and NEISO, will not fulfill the regional frequency response obligation as the regional renewable penetration reaches 80%. Among these three BAs, PJM will have the largest deficiency of frequency response because of the extremely high renewable penetration in some PJM subareas, such as PJM_ROM (100% renewable penetration). Comparatively, TVA and SOCO can still fulfill the BA frequency response obligation when the interconnection-level renewable penetration reaches 80%, mainly because of a large capacity of frequency-responsive hydro power plants.

It is worth noting that the future regional PV penetration and the retirement of synchronous generation may vary with market, technology, politics, environmental factors, some of which are hard to predict precisely. Therefore, for a particular region, the actual compliance of regional frequency response code may differ from the predicted value obtained in this study. The purpose of this study is to deliver some credible case studies based on the developed scenarios. If synchronous generators in a BA cannot fulfill regional frequency response obligation, inertia and governor controls from renewable generation and other resources will be required.

IV. CONCLUSIONS

This paper studied the impact of high PV penetration on regional frequency response. It is found that high PV penetration concentrated in one region will have a large impact on the RoCoF and frequency nadir obtained from the regional frequency. In contrast, the impact on the regional settling frequency is small due to that the settling frequency converges to the system average frequency when the frequency is settled. In addition, the EI regional frequency check found that some regions with high PV penetration can not comply with the frequency response grid code, calling for support of primary frequency response support from other regions with conventional generation governors. This result also indicates a possibility of a primary frequency response market to ensure the system frequency stability while fully leverage the renewable energy resources in different regions.

REFERENCES

1. Commission, F.E.R., *BAL-003-1—Frequency response and frequency bias setting.* Washington, DC, March, 2013.
2. NERC, *Frequency Response Standard Background Document.* 2012.

---

[1] In FRO$_{BA}$ calculation, annual generation and load for each balancing authority are estimated values based on the power flow model.